%Paper: astro-ph/9312026
%From: <fbernard@chipmunk.cita.utoronto.ca>
%Date: Mon, 13 Dec 93 16:17:22 EST

%skewness and kurtosis in large-scale cosmic fields, Plain Tex.
\magnification\magstep 1
\headline={\ifnum\pageno=1\hfil\else\hfil\tenrm--\ \folio\ --\hfil\fi}
\footline={\hfil}
\hsize=6.0truein
\vsize=8.54truein
\hoffset=0.25truein
\voffset=0.25truein
\baselineskip=12pt
\def\init{\tabskip 0pt }
\def\crr{\cr \noalign{\hrule}}

\parskip=0.2 cm
\parindent=1cm
\tolerance=10000
\pretolerance=10000
\def\pp{\parshape 2 0truecm 15truecm 2truecm 13truecm}
\def\apjref#1;#2;#3;#4; {\par\pp#1, { #2}, { #3}, #4 \par}
\def\bookref#1;#2;#3; {\par\pp#1, { #2}, {\rm #3} \par}
\def\prepref#1;#2; {\par\pp#1, { #2} \par}
\def\ie{{\it i.e. }}
\def\eg{{\it e.g.}}
\def\etal{{\it et al. }}
\def\ltsima{$\; \buildrel < \over \sim \;$}
\def\simlt{\lower.5ex\hbox{\ltsima}}
\def\gtsima{$\; \buildrel > \over \sim \;$}
\def\simgt{\lower.5ex\hbox{\gtsima}}
\def\gm{\gamma}
\def\Om{\Omega}

\def\rhob{\overline{\rho}}
\def\mg{\big <}
\def\md{\big >}

\def\mP{{\cal P}}
\def\mQ{{\cal Q}}
\def\mR{{\cal R}}
\def\mU{{\cal U}}

\def\mS{{\cal S}}

\def\d{\hbox{d}}
\def\ii{\hbox{i}}
\def\dta{\delta}

\def\dtag{\delta_{\hbox{galaxy}}}
\def\grad{\nabla}
\def\Mpc{\hbox{Mpc}}

\def\WTH{W_{\hbox{TH}}}
\def\WTHu{\WTH(k\,R_0)}

\def\WTHd{\WTH\left(\vert\vk_1+\vk_2\vert\ R_0\right)}

\def\dWTHd{\WTH'\left(\vert\vk_1+\vk_2\vert\ R_0\right)}

\def\vu{\hbox{\bf u}}
\def\vx{\hbox{\bf x}}
\def\vr{\hbox{\bf r}}

\def\vv{\hbox{\bf v}}
\def\vk{\hbox{\bf k}}

\def\gradx{\nabla_{\hbox{x}}\,}

\def\epsk{\epsilon_{\vk}}
\def\epsku{\epsilon_{\vk_1}}
\def\epskd{\epsilon_{\vk_2}}
\def\epskt{\epsilon_{\vk_3}}

\def\un{^{(1)}}
\def\deux{^{(2)}}
\def\trois{^{(3)}}

\def\Sgal{S^{\hbox{gal}}}

\def\zel{^{\hbox{Zel}}}

\def\intk{\int{\d^3\vk\over(2\pi)^{3/2}}}
\def\intkc{\int{\d^3\vk\over(2\pi)^{3}}}
\def\intkp{\int{\d^3\vk'\over(2\pi)^{3/2}}}
\def\intd{\int{\d^3\vk_1\over(2\pi)^{3/2}}{\d^3\vk_2\over(2\pi)^{3/2}}}
\def\intt{\int{\d^3\vk_1\over(2\pi)^{3/2}}{\d^3\vk_2\over(2\pi)^{3/2}}
{\d^3\vk_3\over(2\pi)^{3/2}}}
\def\intdc{\int{\d^3\vk_1\over(2\pi)^{3}}{\d^3\vk_2\over(2\pi)^{3}}}
\def\inttc{\int{\d^3\vk_1\over(2\pi)^{3}}{\d^3\vk_2\over(2\pi)^{3}}
{\d^3\vk_3\over(2\pi)^{3}}}

\def\put{\mP(\vk_1,\vk_2)}
\def\qut{\mQ(\vk_1,\vk_2)}
\def\pu{1+{\vk_1.\vk_2\over k_1^2}}
\def\qu{1-{\left(\vk_1.\vk_2\right)^2\over k_1^2\,k_2^2}}

\vfill\eject
\headline={\hfil\tenrm--\ \folio\ --\hfil}
\vglue .5 cm
\rightline{CITA 93/44}
\vglue 1.3 truecm
\centerline{\bf SKEWNESS AND KURTOSIS IN}
\centerline{\bf LARGE-SCALE COSMIC FIELDS}
\vskip .4 cm
\centerline{by}
\vskip .4 cm
\centerline{Francis BERNARDEAU}
\vskip .3 cm
\centerline{CITA, 60 St George St., Toronto, Ontario, Canada M5S 1A1}
%\centerline{SPhT, CE Saclay, F-91191 Gif-sur-Yvette, France}
\vskip 1. truecm

\centerline{\bf ABSTRACT}

In this paper, I present the calculation of the
third and fourth moments of both the  distribution function of
the large--scale density and the large--scale divergence of the velocity field,
$\theta$.
These calculations are made by the mean
of perturbative calculations assuming Gaussian initial conditions and
are expected to be valid in the linear or quasi linear regime. The
moments are derived for a top--hat window function and for any cosmological
parameters $\Omega$ and $\Lambda$. It turns out that the
dependence with $\Lambda$ is always very weak whereas the moments
of the
distribution function of the divergence are strongly dependent on $\Omega$.
A method to measure $\Omega$ using the skewness
of this  field has already been presented by Bernardeau et al. (1993).
I show here that the simultaneous measurement of the skewness and the
kurtosis allows
to test the validity of the gravitational
instability  scenario hypothesis. Indeed
there is a combination of the first three moments of $\theta$
that is almost independent of the
cosmological parameters $\Omega$ and $\Lambda$,
$${\left(\mg\theta^4\md-3\mg\theta^2\md^2\right)\ \mg\theta^2\md
\over \mg\theta^3\md^2}\approx 1.5,$$
(the value quoted is valid when the index of the power spectrum at
the filtering scale is close to -1)
so that any cosmic velocity field created by gravitational
instabilities should verify such a property.

\vskip 1 truecm
\centerline {Submitted for publication to ``\sl The Astrophysical Journal''}

 \vfill\eject

\centerline{\bf 1. INTRODUCTION}

At large scales, say for scales larger than 10 $h^{-1}$ Mpc,
the amplitudes of the rms fluctuation observed in the cosmic fields
are still below unity showing that these fields have undergone only a
moderate evolution.
The perturbation theory is then well adapted to address many problems
associated
with these scales. For instance the growing rate of the fluctuations
can be accurately described by the linear approximation
(\eg, Efstathiou \etal 1988). This is a crucial
result for the determination of the initial spectrum. However, the
statistical properties of the large--scale cosmic fields cannot
be reduced to the behavior of the fluctuation spectrum, as it would have been
the case if the fluctuations remain Gaussian. For instance the skewness
observed in the density distribution at such large scales
(see for example Bouchet \etal 1993) is
in agreement with the theoretical predictions based on
calculations using the second order perturbation theory
(Peebles 1980, Goroff \etal 1986, Juszkiewicz, Bouchet \& Colombi 1993a).

The properties of the cosmic velocity field so far have only been considered
in relation with the density field.
In the linear theory,
there is a well
known relationship (Peebles 1980)
between the density and the peculiar velocity field,
$\vu(\vx)=\vv(\vx)-H_0\vr$,
$$\vu(\vx)=-f(\Omega,\Lambda){H_0\over 4\pi}\,\int
{\vx-\vx'\over\vert\vx-\vx'\vert^3}\,\dta(\vx)\,\d^3\vx'\eqno(1)$$
where $f(\Omega,\Lambda)$
is a function depending on the cosmological parameters
and $\delta(\vx)$ is the matter
overdensity field ($\dta(\vx)=\rho(\vx)/\rhob-1$). According to
Peebles (1980), the function $f$ can be accurately approximated by
$f(\Omega)=\Omega^{0.6}$ and Lahav \etal (1991) recently showed that
its $\Lambda$ dependence was extremely weak. So in principle a comparison
of the velocity field with the density field should lead to a determination
of $\Omega$. This analysis has been done using the dipole measurement
(that gives the absolute velocity of our galaxy with respect to the
microwave background) and assuming that
the local density is given by the galaxy field as given by the
1.2 Jy IRAS survey (Strauss \& Davis 1988) or by the QDOT survey
(Rowan-Robinson \etal 1990).
These analysis lead to the conclusion that
$\Omega$ is close to unity (Kaiser \& Lahav 1989).
However this method raised criticisms since
the convergence of the integral in (1) may be quite slow (see Peacock
1992). This difficulty has been partly overcome by Yahil (1991)
who uses a local form of the relation (1),
$$\theta(\vx)\equiv
{1\over H_0}\grad.\vu(\vx)=-f(\Omega,\Lambda)\,\dta(\vx)\eqno(2)$$
to compare the velocity field as determined by Dekel \etal (1991)
with the density field. Using the IRAS galaxy sample, he also concludes
that $\Omega$ should be close to unity. However, Shaya, Tully \& Pierce
(1992) using optical-based study give a slightly lower value for $f$
leading to $\Omega\simlt 0.1$. In any case, we must
have in mind that the light distribution may be a biased indicator
for the actual mass distribution. For small fluctuations the existence of such
a bias can be parametrized by the relationship
$$\dtag(\vx)=b\,\dta(\vx).\eqno(3)$$
As a result the methods previously mentionned only give the value
of the ratio $f(\Omega)/b$. It then would be of great interest to find a
method that would separate these two quantities. Yahil (1991)
proposed a way to do it but he assumed some arbitrary form for
the nonlinear relationship between the galaxy density field and the
matter density field that has no reason to be true.
The nonlinear corrections between the matter density field and the
velocity field
have been investigated either by numerical simulations (Nusser \& Dekel 1991)
or on theoretical basis (Bernardeau 1992b). However these corrections do
not give a way to separate the measurement of $\Omega$ from the bias
effects (see section 4.2 of this paper).

One way to overcome the problem of galaxy biasing is simply
to use only the velocity field as a probe of the large--scale structures.
The linear theory (eq. [1] or equivalently [2]) is obviously of no
help. However some statistical properties of the velocity field
can be derived from perturbative calculations at higher orders.
It is noticeable that the divergence of the peculiar velocity field
is a scalar field
that can be analyzed in a similar way than the density field.
And as data on large--scale velocity field
begin to be available (Lynden-Bell \etal 1988, Dekel, Bertschinger \& Faber
1991) it becomes crucial
to explore the various statistical indicators that could be
of cosmological interests in such a field.
Following the work of Dekel \etal (1991) it seems possible
to build the full 3D velocity field from the mere knowledge of the
line of sight velocities of a sample of galaxies. The reconstruction
is based on the assumption that the large--scale velocity field must
be only potential - non--rotational -  so that it derives from a simple
scalar field. The divergence of the velocity field then contains
as much information as the peculiar velocity field apart from a
possible uniform flow which just corresponds to the invariance of the equations
of motion under a Galilean transformation.

In the following I will assume that the large--scale structures formed by
gravitational instabilities from initial Gaussian fluctuations.
There are indications that the initial fluctuations were actually Gaussian
in the galaxy distribution (see further in this
introduction). Then quantities such
as the skewness, $\mg\dta^3\md$ and
$\mg\theta^3\md$, or the kurtosis,
$\mg\dta^4\md-3\mg\dta^2\md^2$ and
$\mg\theta^4\md-3\mg\theta^2\md^2$,
of the cosmic fields are then of
great interest ($\mg.\md$ is an ensemble average over
the initial fluctuations).
Indeed, because of the general
properties of Gaussian variables, these quantities are zero for the initial
fields, and the use of the linear theory would conclude
that they remain zero.
But in fact gravity, because of the nonlinearities it contains, induces
specific non-Gaussian features
that, even at very large--scales,  will be exhibited in these
parameters in a very precise way. The first example of such a calculation
was given by Peebles (1980) for the skewness. He showed that
its leading term for small $\sigma$ can be derived from a second
order perturbative calculation (and higher orders give only negligible
corrections). Fry (1984) extended these results and showed that
the kurtosis can be obtained from third order calculations.
These results were extended afterwards in several directions. A method
was presented by Bernardeau (1992a) to get the whole series
of the cumulants. The $\Omega$ dependence and the smoothing effects
have been calculated for the density skewness (Bouchet \etal 1992,
Goroff \etal 1986, Juszkiewicz, Bouchet \& Colombi 1993) and a recent paper
(Bernardeau \etal 1993) extends these results to the velocity divergence
presenting the dependence of
its skewness with $\Omega$ and the shape of the power spectrum
for various window functions. The result reads,
$${\mg\theta^3\md\over\mg\theta^2\md^2}\approx {-1\over \Omega^{0.6}}
\,\left[{26\over7}-(3+n)\right],\eqno(4)$$
for a top--hat window function  (volume--weighted
filtering) and a power law spectrum of index $n$.
As can be seen in the relation (4) the skewness
has a dependence with the variance --proportional to its square--
that is specific of Gaussian initial conditions. The generic relation for
non--Gaussian fluctuations would have been
$\mg\theta^3\md\propto \mg\theta^2\md^{3/2}$. Such a property
is also true for the density field and for the same reason. It has
been successfully checked in the IRAS galaxy survey
(Bouchet \etal 1993), bringing strong
support to the Gaussian fluctuations hypothesis.
The relation (4) has been proposed to determine $\Omega$ from
the observed velocity field (Bernardeau \etal 1993).
The purpose of this paper is to extend the result (4) to non-zero values
of $\Lambda$ and to derive a similar property for the kurtosis for
both the density and the velocity fields.
 In part 2, I present the principle of the calculations and
in part 3, the results. They are
restricted to a top--hat window
function due to its special geometrical properties presented
in the appendix. Part 4 is devoted to comments and comparisons with numerical
simulations.

 \vskip 1 cm

\centerline{\bf 2. THE METHOD}
\vskip .5 cm
\leftline{2.1. \sl The initial conditions}

At large scale the divergence of the velocity field,
as the density field, has fluctuations, the distribution of which can
be examined by perturbative calculations. The first order solution
in $\delta(\vx)$ is directly related to the first order in $\theta(\vx)$
through the continuity equation (Eqs. [1-2]).
The variance of this field is then, in the linear regime, driven by the first
order approximation. It reads
$$\mg\theta(\vx)^2\md=f(\Omega)^2\mg\delta(\vx)^2\md\eqno(5)$$
when the growing mode only is taken into account.

In the following, I assume initial Gaussian fluctuations. The density
field, at first order, can then be written
$$
\delta^{(1)}(\vx,t)=D_1(t)\epsilon(\vx)
=D_1(t)\int{\d^3\vk\over (2\pi)^{3/2}}
\,\epsilon_{\vk}\,e^{\ii\vk.\vx}\eqno(6)$$
where $D_1(t)$ if the growing factor (see part 3 for the equation
the solution of which $D_1(t)$ is) and
the random variables $\epsilon_{\vk}$ follow the rules,
$$\eqalign{
\mg\epsilon_{\vk}\epsilon_{\vk'}\md&=\delta_D(\vk+\vk')P(k);\cr
\mg\epsilon_{\vk_1}\dots\epsilon_{\vk_{2p+1}}\md&=0;\cr
\mg\epsilon_{\vk_1}\dots\epsilon_{\vk_{2p}}\md&={1\over 2^pp!}
\sum_{\hbox{permutations}\ (s)}
\mg\epsilon_{\vk_{s_1}}\epsilon_{\vk_{s_2}}\md
\dots\mg\epsilon_{\vk_{s_{2p-1}}}\epsilon_{\vk_{s_{2p}}}\md.\cr}
\eqno(7)$$
where the function $\delta_D$ is the 3D Dirac distribution,
and $P(k)$ is the spectrum of the fluctuations.
The divergence of the velocity field now reads,
$$\theta^{(1)}(\vx,t)=-f(\Omega)
D_1(t)\int{\d^3\vk\over (2\pi)^{3/2}}
\,\epsilon_{\vk}\,e^{\ii\vk.\vx},\eqno(8)$$
at the first order in $\epsilon_{\vk}$. At this order, the fields (6) and (8)
are purely Gaussian and their whole
statistics is determined by the shape of the power spectrum.
Specific features induced by gravity appear when
terms of greater orders are taken into account.

\vskip .5 cm
\leftline{2.2. \sl The equations of the dynamics}

Throughout the paper the matter field is approximated by a
nonrelativistic and self-gravitating fluid at zero pressure.
I do not exclude the existence of a non-zero cosmological constant $\Lambda$.
The expansion factor, $a(t)$, is then solution of the equation,
$${\ddot{a}\over a}=-{4\pi\,G\,\rhob\over 3}+{\Lambda\over 3}\eqno(9)$$
For convenience I will characterize the (time dependent)
cosmological parameters by $\Omega$ and by the
Hubble constant, $H$, and its time derivative. Let me
recall that $\Omega$ is defined by
$$4\pi\,G\,\rhob={3\Omega\over 2}H^2.\eqno(10)$$
(At the present time $H=H_0$.)
The fluid is described by the overdensity field $\dta(\vx,t)$ and
by the peculiar velocity field $\vu(\vx,t)$
assumed to be non-rotational so that
$$\gradx\times\vu(\vx,t)=0.\eqno(11)$$
The equations of the dynamics are then given by (\eg, Peebles 1980),
$$\left.\eqalign{
{\partial \over \partial t}\dta(\vx,t)+{1\over a}\gradx .\left[
(1+\dta(\vx,t))\vu(\vx,t)\right]&=0\cr
{\partial \over \partial t}\vu(\vx,t)+{\dot a \over a}\vu(\vx,t)+{1\over a}
(\vu(\vx,t).\gradx)\vu(\vx,t)&=-{1\over a}\gradx\psi(\vx,t)\cr
\gradx^2\psi(\vx,t)=4\pi G\overline{\rho}a^2\dta(\vx,t),\cr}\right.\eqno(12)$$
where the spatial derivatives are taken with respect to $\vx$.
The first step is to take the divergence of the second equation so that
the system now reads, using the definition of $\Omega$,
$$
\eqalign{
{1\over H}\dot{\dta}(\vx,t)+[1+\dta(\vx,t)]\theta(\vx,t)+{1\over a H}
\gradx .[\dta(\vx,t)\vu(\vx,t)]&=0\cr
\left(2+{\dot{H}\over H^2}\right)\theta(\vx,t)+{1\over H}\dot{\theta}(\vx,t)+
{3\over 2}\Omega\ \dta(\vx,t)+{1\over a H}\gradx .[\vu(\vx,t).\gradx]\vu(\vx,t)
&=0\cr}\eqno(13)$$
where a dot is for a time derivative.

It is convenient to introduce the Fourier transforms of these fields,
$\dta(\vk,t)$ and $\theta(\vk,t)$, defined by
$$\eqalign{
\dta(\vx,t)&=\intk \dta(\vk,t)\exp(\ii \vx.\vk),\cr
\theta(\vx,t)&=\intk \theta(\vk,t)\exp(\ii \vx.\vk).\cr}\eqno(14)$$
Due to the absence of rotational part in the velocity field, $\vu(\vx)$
simply reads
$$\vu(\vx,t)=\intk {-\ii\,\vk\over k^2}\,
\theta(\vk,t)\exp(\ii \vx.\vk).\eqno(15)$$
Using the relations (14-15) I can now write the equation of the
dynamics in the $\vk$-space:
$$\eqalign{
&{1\over H}\dot{\dta}(\vk,t)+\theta(\vk,t)+\intkp
\,\mP(\vk',\vk-\vk')\,\dta(\vk-\vk',t)\,\theta(\vk',t)=0\cr
&
\left(2+{\dot{H}\over H^2}\right)\theta(\vk,t)+{1\over H}\,\dot{\theta}(\vk,t)+
{3\over 2}\,\Omega \dta(\vk,t)=\cr
&-\intkp
\left[\mP(\vk',\vk-\vk')+\mP(\vk-\vk',\vk')-2\mQ(\vk',\vk-\vk')\right]
\theta(\vk-\vk',t)\,\theta(\vk',t),\cr}\eqno(16)$$
where
$$
\eqalignno{
\mP(\vk,\vk')&=1+{\vk.\vk'\over k^2},&(17)\cr
\mQ(\vk,\vk')&=1-{\left(\vk.\vk'\right)^2\over k^2 k'^2}.&(18)\cr}
$$
These functions will be extensively used in the following.
The system (16) is the basic system of equations used for the perturbative
calculations.

\vskip .5 cm
\leftline{2.3. \sl The filtering of the fields}

In practice we have to consider the density and the velocity field after they
have
been filtered by a particular window function. Indeed perturbative
calculations are only valid at (very) large scale, so that the
strong small--scale fluctuations should be smoothed out. I
make the assumption that the existence of nonlinear fluctuations
at small scales will not change the result. This assumption
has been verified for instance for the time dependence of the
variance at large scale in various numerical simulations (see for
instance Efstathiou \etal 1988). There
are no reasons to think that this is not true for the purpose of
these calculations and the comparison with numerical simulations
presented in part 3 confirm the validity of this hypothesis.

For technical reason
I only consider the case of the top--hat window function. However, nothing
in practice prevents to use any other window function. For instance Goroff
\etal (1986) present results for the density field using a Gaussian window
function but in such a case the smoothing corrections have been to be
calculated numerically
for each power spectrum. For a top--hat window function all the results
can be given analytically as shown in the next part.

As it has been noticed by Bernardeau \etal (1993) it is equivalent to apply
the filter to the velocity field or to its divergence: both are linear
operations that commute each other.
Then I focus my interest on the filtered density and
divergence of the velocity field.
I define $\dta(R_0)$ and $\theta(R_0)$ by
$$\eqalignno{
\dta(R_0)&=\int \d^3\vx\ \WTH(\vx)\ \dta(\vx),&(19)\cr
\theta(R_0)&=\int \d^3\vx\ \WTH(\vx)\ \theta(\vx),&(20)\cr
}$$
with
$$\eqalign{
\WTH(\vx)&=1 \hbox{  if  } \vert\vx\vert\le R_0;\cr
\WTH(\vx)&=0 \hbox{  otherwise};\cr}\eqno(21)$$
for a scale $R_0$.
Note that, for the velocity
field, it corresponds to a volume weighted filtering: the velocity is
equally weighted in each point of space, regardless of the local density.
In practice, the velocity is known only at the positions
of the galaxies (or the matter points for a numerical simulation)
so that it is necessary to smooth the field in two steps:
the local field has to be defined in each point
by a small scale (mass-weighted)
filtering and then filtered with a volume weighted procedure. This is
what have been done by Juszkiewicz \etal (1993b).
In (21) the filter is applied at the origin
but the statistics of the filtered fields obviously does not depend
on the position at which the filter is applied.
I also make the hypothesis that a volume average is equivalent
to an ensemble average so that the moments of distribution functions
of $\delta$ and $\theta$ are in principle measurable in a large enough sample.

The quantities (21) can
be expressed with the Fourier transforms of respectively
the density field and the divergence field. They read,
$$\eqalignno{
\dta(R_0)&=\intk \WTH(\vk\,R_0)\,\dta(\vk),&(22)\cr
\theta(R_0)&=\intk \WTH(\vk\,R_0)\,\theta(\vk),&(23)\cr
}$$
where $\WTH(k\,R_0)$ is the Fourier transform of the top--hat window function,
$$\WTH(k\,R_0)={3\over (kR_0)^3} \left(\sin(kR_0)-kR_0 \cos(kR_0)\right).
$$
The variance of these smoothed fields is then given by,
$$\eqalignno{
\sigma^2(R_0)&=\int{\d^3k\over (2\pi)^{3/2}}\,{\d^3k'\over (2\pi)^{3/2}}
\WTH(k\ R_0)\ \mg\dta(\vk)\dta(\vk')\md\ \WTH(k'\ R_0)\cr
&= D_1^2(t)\intkc \WTH^2(k\ R_0)\ P(k),&(24)\cr
\sigma_{\theta}^2(R_0)&
= f^2(\Omega)\ D_1^2(t)\intkc \WTH^2(k\ R_0)\ P(k).&(25)\cr}$$
These results are valid at large scale where, by definition, the linear
approximation can be applied to the determination of the variance of the
density field. The validity domain of the results given in the
following is in principle the same as the validity domain of (24-25).

\vskip .5 cm
\leftline{2.4. \sl Calculation of the skewness}

The principle of this calculation has been presented in previous
papers (Goroff \etal 1986,
Bouchet \etal 1992 for the density field, Bernardeau \etal 1993
for the divergence of the velocity field).
The aim is to calculate the leading order of the third moment
of the distribution functions of
$\dta(R_0)$ or of $\theta(R_0)$. The calculation is based on an
expansion of these
quantities with respect to the random variables $\epsk$ that describe
the initial density and velocity fluctuations. So I write
$$\dta(R_0)=\dta\un(R_0)+\dta\deux(R_0)+\dta\trois(R_0)+\dots\eqno(26)$$
where $\dta\un(R_0)$ is proportional to $\epsk$, $\dta\deux(R_0)$ is
quadratic in $\epsk$ and $\dta\trois(R_0)$ is cubic in $\epsk$...
A similar expansion can be written for $\theta(R_0)$,
$$\theta(R_0)=\theta\un(R_0)+\theta\deux(R_0)+\theta\trois(R_0)+\dots
\eqno(27)$$
Then the ensemble averages of $\left[\dta(R_0)\right]^3$ and
$\left[\theta(R_0)\right]^3$ read
$$\eqalignno{
\mg\dta(R_0)^3\md&=
\mg\left[\dta\un(R_0)+\dta\deux(R_0)+\dta\trois(R_0)+\dots\right]^3\md\cr
&=\mg\left[\dta\un(R_0)\right]^3\md
+3\,\mg\left[\dta\un(R_0)\right]^2\dta\deux(R_0)\md+\dots,&(28)\cr
\mg\theta(R_0)^3\md&=
\mg\left[\theta\un(R_0)+\theta\deux(R_0)+\theta\trois(R_0)+\dots\right]^3\md\cr
&=\mg\left[\theta\un(R_0)\right]^3\md
+3\,\mg\left[\theta\un(R_0)\right]^2\theta\deux(R_0)\md+\dots&(29)\cr}
$$
The first terms of these expansions are zero due to the general properties
of Gaussian variables. The leading terms will be respectively
$3\mg\left[\dta\un(R_0)\right]^2\dta\deux(R_0)\md$
and
$3\mg\left[\theta\un(R_0)\right]^2\theta\deux(R_0)\md$
that involve the product of four Gaussian variables. Then the ratios,
$$\eqalignno{
S_3&={\mg\dta(R_0)^3\md\over\mg\dta(R_0)^2\md^2},&(30)\cr
S_{3\theta}&={\mg\theta(R_0)^3\md\over\mg\theta(R_0)^2\md^2},&(31)\cr}
$$
are finite at large scale. The purpose of the next part is to calculate
these
limits $S_3$ and $S_{3\theta}$ as a function of the cosmological parameters
$\Omega$ and $\Lambda$ and the shape of the power spectrum.

\vskip .5 cm
\leftline{2.5. \sl Calculation of the kurtosis}

The skewness is a sort of measure of the asymmetry of the distribution
function. The kurtosis measures the flatness of the distribution function
compared to what would be expected from a Gaussian distribution.
The leading order of this term involves both the
second order and third order of the dynamics. It reads
$$\eqalign{
\mg\dta^4(R_0)\md&-3\mg\dta^2(R_0)\md^2\approx
\ 6\,\mg\left[\dta\un(R_0)\right]^2\left[\dta\deux(R_0)\right]^2\md
+4\,\mg\left[\dta\un(R_0)\right]^3\dta\trois(R_0)\md\cr
&-6\,\mg\left[\dta\un(R_0)\right]^2\md\ \mg\left[\dta\deux(R_0)\right]^2\md
-12\,\mg\left[\dta\un(R_0)\right]^2\md\
\mg\dta\un(R_0)\dta\trois(R_0)\md\cr}\eqno(32)$$
for the density field and
$$\eqalign{
\mg\theta^4(R_0)\md&-3\mg\theta^2(R_0)\md^2\approx
\ 6\,\mg\left[\theta\un(R_0)\right]^2\left[\theta\deux(R_0)\right]^2\md
+4\,\mg\left[\theta\un(R_0)\right]^3\theta\trois(R_0)\md\cr
&-6\,\mg\left[\theta\un(R_0)\right]^2\md\ \mg\left[\theta\deux(R_0)\right]^2\md
-12\,\mg\left[\theta\un(R_0)\right]^2\md\
\mg\theta\un(R_0)\theta\trois(R_0)\md\cr}\eqno(33)$$
for the divergence of the velocity field.
In the calculation of these differences, we can notice that
the terms that are subtracted actually appear in the expression
of the fourth moment. They contain the non--connected part of the moment
(so called since they factorize in products of moments) and will exactly vanish
when the fourth order cumulant is calculated.
For instance the fourth cumulant of the
density formally reads
$$\mg\dta^4\md_c
\approx 6\,\mg\left[\dta\un(R_0)\right]^2\left[\dta\deux(R_0)\right]^2\md_c+
4\,\mg\left[\dta\un(R_0)\right]^3\dta\trois(R_0)\md_c,$$
where the subscript $c$ means that only the connected terms are retained.
We call then $S_4$ and $S_{4\theta}$ the large--scale limits of the ratios,
$$\eqalignno{
S_4&={\mg\dta^4\md-3\mg\dta^2\md^2\over\mg\dta^2\md^3},&(34)\cr
S_{4\theta}&={\mg\theta^4\md-3\mg\theta^2\md^2\over\mg\theta^2\md^3}.&(35)\cr}
$$

The next part is devoted to the derivation of the expressions of
$S_3$, $S_{3\theta}$, $S_4$ and $S_{4\theta}$.

 \vskip 1 cm

\centerline{\bf 3. THE CALCULATIONS}

\vskip .5 cm
\leftline{3.1. \sl The dynamics of the spherical collapse}

To calculate the previous limits, the dynamics
of the spherical collapse is quite helpful. In general the
equation describing the evolution of
the size of spherical symmetric perturbation reads,
$$\ddot{R} =-{G\,M(<R)\over R^2}+{\Lambda\over 3}R,\eqno(36)$$
where $R$ is the size of the perturbation and $M(<R)$ is the constant mass
within the radius $R$. The dot stands for a time derivative.
This equation can be written in term of the overdensity $\dta$
defined by,
$\dta=\left({R/R_0}\right)^{-3}-1$,
where $R_0$ is the initial (comoving) radius of the perturbation.
The equation (36), is equivalent to (using Eq. [10])
$$-{1\over2}H^2\Omega\,{1\over 1+\dta}-{2\over3}{H\dot{\dta}\over(1+\dta)^2}
-{1\over3}{\ddot{\dta}\over(1+\dta)^2}
+{4\over 9} {\dot{\dta}^2\over (1+\dta)^3}
=-{1\over2}H^2\Omega.\eqno(37)$$
The latter equation can be linearized in $\dta$.
Let me write this linear term as
$$\dta\un(t)=D_1(t)\dta_i$$
where $\dta_i$ gives the strength of the initial fluctuation
and is supposed to be small. The function $D_1(t)$ is the time
dependence of the growing mode and
it is the solution of the equation,
$$\ddot{D_1}+2H\,\dot{D_1}-{3\over2}H^2\,\Omega\,D_1=0,\eqno(38)$$
which is a growing function of time.
The function $D_1$
is proportional to the expansion factor when $t$ is small or
for an Einstein--de Sitter universe,
but this is not true in general. The second order in
$\dta$ (when $\dta_i$ is assumed to be the small parameter) reads
$$\dta\deux(t)=D_2(t)\,{\dta_i^2\over 2}$$
where the function $D_2(t)$ is solution of the equation,
$$\ddot{D_2}+2H\,\dot{D_2}-{3\over2}H^2\,\Omega\,D_2={3}
H^2\,\Omega\,D_1^2+{8\over3}\dot{D_1}^2,\eqno(39a)$$
and verifies
$$D_2(t)\sim {34\over 21} D_1^2(t)\ \ \hbox{when}\ \ t\to0.\eqno(39b)$$
Similarly the third order reads
$$\dta\trois(t)=D_3(t)\,{\dta_i^3\over 6}$$
with
$$\ddot{D_3}+2H\,\dot{D_3}-{3\over2}H^2\,\Omega\,D_3={9}
H^2\,\Omega\,D_1\,D_2+{8}\dot{D_1}\,\dot{D_2}
-{8}D_1\,\dot{D_1}^2\eqno(40a)$$
and
$$D_3(t)\sim {682\over 189} D_1^3(t)\ \ \hbox{when}\ \ t\to0.\eqno(40b)$$
In the following I will use the functions $D_1, D_2$ and $D_3$ and
their time derivatives to build the second and third order of the
density and velocity fields. There are two other functions that are convenient
to consider. They are given by the evolution of the divergence
of the velocity field, given by $\theta(t)={3\dot{R}/ HR-3}$
all along the  collapse. This function can be expanded
with respect to $\dta_i$ in the form,
$$\theta(t)=-f(\Omega,\Lambda)\left[
E_1(t)\dta_i+E_2(t){\dta_i^2\over 2}+E_3(t){\dta_i^3\over 6}
+\dots\right],\eqno(41)$$
Where the factor $f(\Omega,\Lambda)$ is given by definition by
$$f(\Omega,\Lambda)={a\over D_1}{\d D_1\over \d a}.\eqno(42)$$
{}From the equation of the dynamics we get
$$\eqalignno{
E_1(t)&=D_1(t);\cr
E_2(t)&=D_1{\d\over \d D_1}\left[D_2-D_1^2\right];&(43)\cr
E_3(t)&=D_1{\d\over \d D_1}\left[D_3-3 D_2\,D_1+2 D_1^3\right].&(44)\cr}
$$
The exact analytic form of these functions can be obtained from
the solution of the spherical collapse (\eg, Peebles 1980), but they
are complicated and it is unnecessary to give them explicitly.
For a non-zero $\Lambda$ they have to be integrated numerically from
the equations (39-40). The time variation of the ratios $D_2(t)/D_1(t)$,
$D_3(t)/D_1(t)^3$,... appears simply as a variation of these
ratios with the cosmological parameters $(\Omega,\Lambda)$ as they
cover a given trajectory in the $(\Omega,\Lambda)$ plan. Here, I present
the results of numerical integrations only in the case of zero curvature,
$\Omega+\Lambda/3H^2=1$ and the analytical results for $\Lambda=0$. The
ratios are given as a function of $\Omega$ in Figs. 1-2.

\vskip .5 cm
\leftline{3.2. \sl Expressions of $\dta\un(R_0)$ and $\theta\un(R_0)$}

The fields at the first order are given by the linear solution (Eqs. [6, 8]).
As a result we have
$$\dta\un(R_0)=\intk\epsk\,\WTHu\,D_1(t)\eqno(45)$$
and
$$\theta\un(R_0)=-f(\Omega,\Lambda)\intk\epsk\,\WTHu\,D_1(t).\eqno(46)$$

\vskip .5 cm
\leftline{3.3. \sl Expressions of $\dta\deux(R_0)$ and $\theta\deux(R_0)$}

We have to write the system (18) at the second order in $\epsk$.
It leads to the system,
$$\eqalign{
&{1\over H}\dot{\dta}\deux(\vk,t)+\theta\deux(\vk,t)=a{\d D_1\over \d a}D_1
\intkp
\mP(\vk',\vk-\vk')\epsilon_{\vk-\vk'}\,\epsilon_{\vk'}
\cr
&\left(2+{\dot{H}\over H^2}\right)\theta\deux(\vk,t)
+{1\over H}\,\dot{\theta}\deux(\vk,t)+
{3\over 2}\,\Omega \dta\deux(\vk,t)=\cr
&-\left[a{\d D_1\over \d a}\right]^2\intkp
\left[\mP(\vk',\vk-\vk')+\mP(\vk-\vk',\vk')-2\mQ(\vk',\vk-\vk')\right]
\epsilon_{\vk-\vk'}\,\epsilon_{\vk'}\cr}\eqno(47)$$

It is convenient to define the notations $\mP_{i,j}$ and $\mQ_{i,j}$ by
$$\mP_{i,j}=\mP(\vk_i,\vk_j),\ \ \ \mQ_{i,j}=\mQ(\vk_i,\vk_j).$$
In the following the indices $(i,j)$ will run between 1 and 3.
As a result of Eq. (47) we obtain
$$\eqalign{
\dta\deux(R_0)=&\intd\epsku\epskd\,\WTHd\cr
&\times\left(D_1^2\left[\mP_{1,2}-{3\over2}\mQ_{1,2}\right]
+{3\over2}\,D_2\,\mQ_{1,2}\right)\cr}\eqno(48)$$
and
$$\eqalign{
\theta\deux(R_0)=&-\intd\epsku\epskd\WTHd\cr
&\times f(\Omega,\Lambda)
\left(D_1^2\left[\mP_{1,2}-{3\over2}\mQ_{1,2}\right]
+{3\over4}\,E_2\,\mQ_{1,2}\right).\cr}\eqno(49)$$
These results have already been obtained by Bouchet \etal (1992) for the
density field and by Bernardeau \etal (1993) for the velocity field
in case of $\Lambda=0$.

\vskip .5 cm
\leftline{3.4. \sl Expressions of $\dta\trois(R_0)$ and $\theta\trois(R_0)$}

The calculation of these functions is based on the same principle as
for the second order. The system (16) has now to be written at the third
order in $\epsk$,
$$\eqalign{
&\ \ \ \ {1\over H}\dot{\dta}\trois(\vk,t)+\theta\trois(\vk,t)=\cr
&-\intkp
\mP(\vk',\vk-\vk')\left[\dta\un(\vk-\vk')\,\theta\deux(\vk')+
\dta\deux(\vk-\vk')\,\theta\un(\vk')\right]
\cr
&\ \ \ \ \left(2+{\dot{H}\over H^2}\right)\theta\trois(\vk,t)
+{1\over H}\,\dot{\theta}\trois(\vk,t)+
{3\over 2}\,\Omega \dta\trois(\vk,t)=\cr
&-\intkp \left[
\mP(\vk',\vk-\vk')+\mP(\vk-\vk',\vk')-2\,\mQ(\vk',\vk-\vk')\right]
\,\theta\un(\vk-\vk')\,\theta\deux(\vk')\cr}\eqno(50)$$

I have to define an extra function that will be useful for the intermediate
calculations, $F_3(t)$, defined by the equation,
$$\ddot{F_3}+2H\,\dot{F_3}-{3\over2}H^2\,\Omega\,F_3={3\over2}
H^2\,\Omega\,D_1^3\eqno(51)$$
and which behaves like $9\,D_1^3(t)/10$ when $t$ is small. This function will
not enter in the expressions of the kurtosis so this equation has not
to be solved.
Then we can search a solution of the system (50) having the form,
$$\eqalign{
\dta\trois(\vk)=(2\pi)^{3/2}&
\intt\epsku\epskd\epskt\,\dta_{D}(\vk-\vk_1-\vk_2-\vk_3)\cr
&\times\left[
D_1^3\,\mR_{1}(\vk_1,\vk_2,\vk_3)+D_1\,D_2\,
\mR_{2}(\vk_1,\vk_2,\vk_3)\right.\cr
&\ \ \ \ \ \ \left.
+D_3\,\mR_{3}(\vk_1,\vk_2,\vk_3)+F_3\,\mR_{4}(\vk_1,\vk_2,\vk_3)\right]\cr
}\eqno(52)$$
$$\eqalign{
\theta\trois(\vk)=(2\pi)^{3/2}&
\intt\epsku\epskd\epskt\,\dta_{D}(\vk-\vk_1-\vk_2-\vk_3)\cr
&\times\left[
a{\d D_1\over \d a}\,
D_1^2\,\mS_{1}(\vk_1,\vk_2,\vk_3)+a{\d D_1\over \d a}
\,D_2\,\mS_{2}(\vk_1,\vk_2,\vk_3)\right.\cr
&\ \ \ \ \ \left.+a{\d D_2\over \d a} D_1
\mS_{3}(\vk_1,\vk_2,\vk_3)
+a{\d D_3\over \d a}
\,\mS_{4}(\vk_1,\vk_2,\vk_3)\right.\cr
&\ \ \ \ \ \left.+a{\d F_3\over \d a}
\,\mS_{5}(\vk_1,\vk_2,\vk_3)\right]\cr}\eqno(53)$$
where the functions $\mR_{1}\dots\mS_{5}$ are homegeneous functions
of the wave vectors $(\vk_1,\vk_2,\vk_3)$ and have to satisfy
the system (50). From the properties of the
functions $D_1$, $D_2$, $D_3$ and $F_3$ (Eqs. [38, 39, 40, 51]),
one can find a natural solution of this system. Its writing is a bit
complicated so that I introduce new simplified notations,
$$
\eqalign{
\mP_{ij,k}&=\mP(\vk_i+\vk_j,\vk_k),\ \ \ \
\mP_{i,jk}=\mP(\vk_i,\vk_j+\vk_k).\cr}$$
and similar definitions for $\mQ_{ij,k}$ and $\mQ_{i,jk}$.
Note that $\mQ_{ij,k}=\mQ_{k,ij}$ but in general $\mP_{ij,k}\ne\mP_{k,ij}$.
The homogeneous functions $\mR_1,\dots,\mS_5$ then read
$$\eqalign{
\mR_{1}&=
\left({1\over2}\mP_{3,12}+{1\over2}\mP_{12,3}-{1\over3}\mQ_{3,12}
\right)\mP_{1,2}
+\left(-{3\over2}\mP_{12,3}-{4\over3}\mP_{3,12}
+{5\over2}\mQ_{3,12}\right)\,\mQ_{1,2};\cr
\mR_{2}&=
{3\over4}\left[\mP_{3,12}+\mP_{12,3}-3\mQ_{3,12}\right]\mQ_{1,2};\cr
\mR_{3}&=
{3\over8}\mQ_{3,12}\,\mQ_{1,2};\cr
\mR_{4}&=
{2\over3}\mQ_{3,12}\,\mP_{1,2}
-\left({1\over3}\mP_{3,12}+{1\over2}\mQ_{3,12}\right)\mQ_{1,2};\cr
\mS_{1}&=
\left(-{1\over2}\mP_{3,12}-{1\over2}\mP_{12,3}+\mQ_{3,12}\right)\mP_{1,2}
+\left({5\over2}\mP_{12,3}+{3\over2}
\mP_{3,12}-{15\over2}\mQ_{3,12}\right)\,\mQ_{1,2};\cr
\mS_{2}&=
{3\over4}\left(-\mP_{12,3}+3\,\mQ_{3,12}\right)\mQ_{1,2};\cr
\mS_{3}&=
{3\over4}\left(-\mP_{3,12}+3\,\mQ_{3,12}\right)\mQ_{1,2};\cr
\mS_{4}&=-\mR_{3};\cr
\mS_{5}&=-\mR_{4}.\cr}$$
So we now have the expression of $\dta\trois(R_0)$ and $\theta\trois(R_0)$
(using eventually the functions $E_2$ and $E_3$ for the latter).
These expressions seem rather complicated but further simplications
are unnecessary. The functions $\mP$ and $\mQ$ defined in (17-18) are
well adapted for a convolution with
a top--hat window function as shown in the appendix.

\vskip .5 cm
\leftline{3.5. \sl Expressions of the skewness and the kurtosis}

The expressions
(28) and (32) are then calculated using the general property (7) of
Gaussian variables. It leads to the expressions
for $S_3$ and $S_4$,
$$\eqalign{
S_3(R_0)=&{3\over \sigma^4(R_0)}
\intdc P(k_1)\,P(k_2)\,W_1\,W_2\,W_{12}\,\mU_{1,2}\cr}\eqno(54)$$
and
$$\eqalign{
S_4(R_0)={1\over \sigma^6(R_0)}
&\Bigg[6\inttc P(k_1)\,P(k_2)\,P(k_3)\,W_1\,W_{12}\,W_{23}\,W_3\cr
&\times\left[\mU_{1,2}\mU_{2,3}+\mU_{1,2}\mU_{3,2}+\mU_{2,1}\mU_{2,3}+
\mU_{2,1}\mU_{3,2}\right]\cr
&+4\inttc P(k_1)\,P(k_2)\,P(k_3)\,W_{123}\,W_1\,W_2\,W_3\cr
&\times\left[
D_1^3\,\mR_{1}+D_1\,D_2\,\mR_{2}+D_3\,\mR_{3}+F_3\,\mR_{4}\right]
\Bigg],\cr}\eqno(55)$$
where
$$\eqalign{
\mU_{i,j}&=D_1^2\left[\mP_{i,j}-{3\over2}\mQ_{i,j}\right]
+{3\over2}\,D_2\,\mQ_{i,j},\cr
W_i&=\WTH(k_i\,R_0),\
W_{ij}=\WTH\left(\vert\vk_i+\vk_j\vert\ R_0\right),\ \cr
W_{ijk}&=\WTH\left(\vert\vk_i+\vk_j+\vk_k\vert\ R_0\right).\cr}$$
Similar expressions can be obtained for the velocity field.
The previous calculations lead to technical results that fortunately reduce
to very simple expressions.
Indeed the values of the integrals appearing in these expressions
are all given at the end of the appendix and are very simple.
Careful summations of these terms (that must take into account all
symmetry factors) lead to the following results,
$$\eqalignno{
S_3(R_0)&=\left[3{ D_2\over D_1^2}+\gm_1\right];&(56)\cr
S_{3\theta}(R_0)&
={-1\over f(\Omega,\Lambda)}\left[3{ E_2\over D_1^2}+\gm_{1\theta}\right]
;&(57)\cr
S_4(R_0)&=4{D_3\over D_1^3}+12{D_2^2\over D_1^4}
+\left(14{D_2\over D_1^2}-2\right)\gm_1+{7\over3}\gm_1^2
+{2\over3}\gm_2;&(58)\cr
S_{4\theta}(R_0)&={1\over f^2(\Omega,\Lambda)}
\left[4{E_3\over D_1^3}+12{E_2^2\over D_1^4}
+\left(12{E_2\over D_1^2}+2{D_2\over D_1^2}-2\right)\gm_{1\theta}+{7\over3}
\gm_{1\theta}^2
+{2\over3}\gm_{2\theta}\right],&(59)\cr}$$
where
$$\eqalign{
\gm_1&=
{\d\log[\sigma^2(R_0)]\over \d\log R_0},\ \
\gm_2={\d^2\log[\sigma^2(R_0)]\over \d\log^2 R_0},\cr
\gm_{1\theta}&={\d\log[\sigma_{\theta}^2(R_0)]\over \d\log R_0},\ \
\gm_{2\theta}={\d^2\log[\sigma_{\theta}^2(R_0)]\over \d\log^2 R_0}
.\cr}$$

In all cases only the leading term has been given. The
expected corrections,
due to higher order terms, contain an extra factor $\sigma^2(R_0)$
(or $\sigma_{\theta}^2[R_0]$) so
they are expected to be negligible at large scale.

In Figs. (1-2) it is shown that the $\Omega$ and $\Lambda$ dependences of
$D_2/D_1^2$, $D_3/D_1^3$, $E_2/D_1^2$ and $E_3/D_1^3$ are extremely weak
and, for the values of $\Omega$ and $\Lambda$ of interest, can be
safely approximated by their value when $\Omega=1$, $\Lambda=0$.
So we end up with
$$\eqalignno{
&S_3(R_0)=
{34\over7}+\gm_1;&(60)\cr
&S_{3\theta}(R_0)=
{-1\over \Omega^{0.6}}\left[{26\over7}+\gm_{1\theta}\right];&(61)\cr
&S_4(R_0)=
{60712\over 1323}+{62\over3}\gm_1+{7\over3}\gm_1^2
+{2\over3}\gm_2;&(62)\cr
&S_{4\theta}(R_0)=
{1\over \Omega^{1.2}}\left[
{12088\over 441}+{338\over21}\gm_{1\theta}+{7\over3}\gm_{1\theta}^2
+{2\over3}\gm_{2\theta}\right].&(63)\cr}$$

One can notice that the effects of smoothing introduce a dependence of
these coefficients with the scale through the successive logarithmic
 derivatives
of the variance. If the smoothing were not taken into account
one would have obtained $S_3=34/7$ and $S_4=60712/1323$ (\eg, Bernardeau
1992) and it corresponds to a scale independent $\sigma$, not necessarily
to a small smoothing radius limit. Moreover {34/7} does not appear
as an upper bound for the skewness of the density field. The smoothing
actually mixes scales and for a variance decreasing with scale it lowers
both the skewness and the kurtosis. However for particular initial
power spectra (such the ones given by Hot Dark Matter models) the skewness
may take values greater than the one obtained without smoothing. Such
a trend is indeed observed in numerical simulations (Bouchet \& Hernquist,
1992).

\vskip .5 cm
\leftline{3.6. \sl Power law spectrum and scale dependence}

The scale dependences that appear in the expressions (60-63)
are only due
to a possible change of the shape of the power spectrum with scale. A self
similar power spectrum, \ie, a power law of $k$,
$$P(k)\propto k^{n},\eqno(64)$$
leads to values of
$S_3$, $S_{3\theta}$, $S_4$ and $S_{4\theta}$ that are independent
of scale (as long as the quasilinear regime is valid).
The previous results are then very simple since $\gm_2=\gm_{2\theta}=0$,
and read
$$\eqalignno{
S_3&={34\over7}-(n+3);&(65)\cr
S_{3\theta}&={-1\over \Omega^{0.6}}\left[
{26\over7}-(n+3)\right];&(66)\cr
S_4&={60712\over 1323}-{62\over3}(n+3)+
{7\over3}(n+3)^2;&(67)\cr
S_{4\theta}&={1\over \Omega^{1.2}}\left[
\ \ {12088\over 441}-{338\over21}(n+3)+{7\over3}(n+3)^2\right].&(68)\cr}
$$
In such a case the results for the skewness have already been given in previous
papers (Juszkiewicz \etal 1993, Bernardeau \etal 1993).

These functions are given in Figs. 3-5 as a function
of $\Omega$ when $\Lambda=0$ and when $\Omega+\Lambda/3H^2=1$ and
for different values of the index of power law spectra.
In Fig. 6, I present these functions as a function of the
power law index $n$.
In practice it is very useful to approximate the actual scale dependence
of the variance of the density field by a power law,
$$\sigma^2(R_0)\propto R_0^{-(3+n)}\eqno(69)$$
where
$$n\equiv -{\d\log \sigma^2(R_0)\over \d\log R_0}-3.\eqno(70)$$
With such an approximation one gets the relationships (65-68).
For the spectra of
interest it turns out to be a good approximation (see Figs. 7-8).
In these figures I give the expected behavior of the skewness and the kurtosis
as a function of scale using eqs. (60-63) (thick lines). I also
give the kurtosis using the approximation (67) and the
relations (67, 68) (thin lines).
The decrease of the skewness and the kurtosis is due
to the variation of the index $n$ with scale (upper panel).
This variation is more important
for the CDM spectrum (Fig. 7) than for the spectrum proposed by Peacock
(1991) (Fig. 8)
to fit the observation in the galaxy distribution. It corresponds
to the well known discrepancy between the shape of the CDM spectrum and the
one derived from the actual galaxy clustering properties.

In table 1, I summarize the results of these perturbative calculations for the
skewness and the kurtosis for an Einstein--de Sitter Universe.
The parameters obtained in this paper
for the density field are somehow lower than the ones obtained by
Goroff \etal (1986) but
this is due to the fact that they used a Gaussian window function
instead of a top--hat window function.
This effect is particularly
important when the dependence of $\sigma$ with the scale is strong.
However, the dependence of the results with the cosmological parameters,
$\Omega$ and $\Lambda$, which has not been considered by Goroff \etal,
is roughly independent of the shape of the window function.

\midinsert
{\parindent 2 em
\narrower\sl\baselineskip 12pt
\noindent {\bf Table 1:} Predictions of the perturbation theory for
$S_3$, $S_4$, $S_{3\theta}$ and $S_{4\theta}$ vs. spectral slope
for an Einstein--de Sitter Universe.
$$
\baselineskip=11pt\vbox{\init\halign to 6.0 truein{
\strut#&#\tabskip=.5em plus .5em&
\hfil$#$\hfil&
$\,$#&
\hfil$#$\hfil&
$\,$#&
\hfil$#$\hfil&
$\,$#&
\hfil$#$\hfil&
$\,$#&
\hfil$#$\hfil&
$\,$#&
\hfil$#$\hfil&
#\tabskip 0pt\crr
&&n&&S_3&&S_4&&S_{3\theta}&&S_{4\theta}&&S_{4\theta}/S_{3\theta}^2&\crr
&&\ &&\ &&\ &&\ &&\ &\cr
&&-3&&{34/7}&&{{60712}/ {1323}}&&{{26}/ 7}&&{{12088}/ {441}}&&
{{3022}/ {1521}}&\cr
&&-2&&{{27}/ 7}&&{{36457}/ {1323}}&&{{19}/ 7}&&{{6019}/ {441}}&&
{{6019}/ {3249}}&\cr
&&-1&&{{20}/ 7}&&{{18376}/ {1323}}&&{{12}/ 7}&&{{2008}/ {441}}&&
{{251}/ {162}}&\cr
&&0&&{{13}/ 7}&&{{6469}/ {1323}}&&{5/ 7}&&{{55}/ {441}}&&{{11}/ {45}}&\cr
&&1&&{6/ 7}&&{{736}/ {1323}}&&-{2/ 7}&&{{160}/ {441}}&&-&\cr
&&\ &&\ &&\ &&\ &&\ &\crr}}\baselineskip=19pt$$
\par}
\endinsert

\vskip .5 cm
\leftline{3.7. \sl Validity domain and comparison with numerical simulations}

The results that are given throughout the paper are thought to be
valid in the linear or quasi-linear regime. Numerical simulations
have been done to check the evolution of the variance compared to what
is given by perturbative calculations and the agreement with the
predictions based on the linear approximation hold for $\sigma^2$ up
to 0.8 at a 15\% precision level (\eg, Efstathiou \etal 1988 or
Hamilton \etal 1991).
In principle the results given in this paper are valid in the same domain.
Indeed,
results of numerical simulations show that the agreement for $S_3$ and $S_4$
(Bouchet \& Hernquist 1992)
are extremely good even for values of $\sigma$ close to unity.
Juszkiewicz \etal (1993) extend this analysis to the behavior
of $S_{3\theta}$ for a Gaussian filter showing a good agreement
with the theoretical predictions.
I present here a new set of results for the skewness and the kurtosis
for the density field using a top--hat filter to check, in particular,
the accuracy of the scale dependence of the theoretical results for
a CDM power spectrum.
The measures have been made in an adaptive P$^3$M
simulation, kindly provided by Couchman (Couchman 1991),
with CDM initial conditions, $H_0=50\ $ km s$^{-1}$ Mpc$^{-1}$ and
$\Omega=1.$ It is done in a cubic box of side 400 Mpc with
periodic boundary conditions and contains $2.1\ 10^6$ particles
and at the end of the simulation the amplitude of the density fluctuations
is $\sigma_8=0.97$. The resulting
field has been filtered with a top--hat window function at four different
radius and at two different timesteps corresponding to an
expansion factor of 0.6 and 1 (in units of the final expansion
factor). This procedure allows
to check the stability of the result for different values
of $\sigma$. For each radius it is possible to calculate the index
$n$ of the power spectrum (Eq. [69]) as well as the
parameter $\gamma_2\ (\equiv -\d n/\d\log(R_0))$ and then the expected values
of $S_3$ and $S_4$. These numbers are given in respectively the second to the
fourth column of table 2. In parentheses is also given the value of
$S_4$ when the $\gamma_2$ term has been dropped.
The remaining columns correspond to the measurement
of these quantities together with
$\sigma$ at the two different timesteps.

The density has been measured in spherical cells
of given radius centered on the 50$^3$ points of a grid.
The actual second, third and fourth moments of the density distributions
thus obtained
have been calculated, as well as the ratios $S_3$ and $S_4$.
The estimations
of the errors have been made by taking the extreme deviations
that are obtained when the volume of the simulation is divided
in eight different
parts. It provides an estimation of the sampling errors for
the measurement of the skewness and the kurtosis expected
in a cubic
box of size $100 h^{-1}$Mpc ($h=H_0/100\ $km s$^{-1}$ Mpc$^{-1}$)
which is the order of magnitude of the size of
available catalogues. As it can be seen the measurements are in
good agreement with the theoretical predictions even for values of
$\sigma$ slightly higher than unity.
Unfortunately the determination of the same quantities
for the velocity field reveals, for a top--hat filter, to be more
complicated and there are no available results.
The results for the density field are however quite encouraging since
if they are transposed to the divergence of the velocity field,
it seems possible to measure the skewness and the kurtosis with enough
accuracy to put strong constraint on $\Omega$, for filtering radius
up to 15 $h^{-1}$ Mpc. At such a scale
the perturbative calculations should be valid (and for the density field
the analytical results are valid down to 5 $h^{-1}\Mpc$).

\midinsert
{\parindent 2 em
\narrower\sl\baselineskip 12pt
\noindent {\bf Table 2:} Comparison of the theoretical predictions with
numerical simulations for a CDM initial spectrum, $\Omega=1$ and
$H_0=50\ $km s$^{-1}$ Mpc$^{-1}$. In parentheses is given the value of
$S_4$ when the $\gamma_2$ parameter is neglected, eq. (67).
$$
\baselineskip=11pt\vbox{\init\halign to 6.0 truein{
\strut#&#\tabskip=.5em plus .5em&
\hfil$#$\hfil&
$\,$#&
\hfil$#$\hfil&
$\,$#&
\hfil$#$\hfil&
$\,$#&
\hfil$#$\hfil&
#\tabskip 0pt\crr
&& &&\hbox{predictions}&&\hbox{measures}\ a/a_0\approx0.6&&a/a_0=1&\crr
&&R_0/\Mpc\ \ \ \ n\ \ \ \ \ \ \gamma_2&&S_3\ \ \ \ S_4\ \ \phantom{(S_4)}
&&\sigma\ \ \ \ \ \ S_3\ \ \ \ \ \ S_4
&&\sigma\ \ \ \ \ \ S_3\ \ \ \ \ \ S_4\crr
&&\ &&\ &&\ &\cr
&&10\ \ -1.34\ \ -0.46&&3.2\ \ \ 17.7\ (18.1)
&&0.92\ \ 3.2^{+0.3}_{-0.3}\ \
18^{+5\phantom{0}}_{-5}&&1.52\ \ 3.4^{+0.5}_{-0.5}\ \
20^{+5\phantom{0}}_{-8}&\cr
&&20\ \ -0.98\ \ -0.57&&2.9\ \ \ 13.3\ (13.7)
&&0.46\ \ 2.8^{+0.4}_{-0.4}\ \
13^{+5\phantom{0}}_{-5}&&0.74\ \ 2.8^{+0.5}_{-0.6}\ \
13^{+6\phantom{0}}_{-7}&\cr
&&30\ \ -0.73\ \ -0.64&&2.6\ \ \ 10.6\ (11.0)
&&0.29\ \ 2.4^{+1.1}_{-0.8}\ \
11^{+20}_{-7}&&0.47\ \ 2.7^{+0.7}_{-0.9}\ \ 13^{+10}_{-12}&\cr
&&40\ \ -0.54\ \ -0.71
&&2.4\ \ \ \phantom{0}8.7\ \phantom{0}(9.2)&&0.21\ \ 2.2^{+1.5}_{-1}\ \
16^{+23}_{-20}&&0.33\ \ 2.6^{+0.6}_{-1.6}\ \ 15^{+14}_{-16}&\cr
&&\ &&\ &&\ &\crr}}\baselineskip=19pt$$

\par}
\endinsert

A lot of numerical works still remain to be done to check the various
features expected for the skewness and the kurtosis with regards to the
cosmological parameters. Moreover it is not clear whether the validity
domain is the same for any power spectrum and whether the errors
associated with the finite size of the sample depend only on the
local index of the power spectrum or on its global shape.
In the latter case the sampling errors would have to be
determined by numerical simulations for any model
to be tested against the observations.

 \vskip 1 cm

\centerline {\bf 4. COMMENTS}

\vskip .5 cm
\leftline{4.1. \sl Higher order moments}

Obviously the skewness and the kurtosis do not carry all the information
available from the distribution of the divergence of the
velocity field. The (connected part of the) moments of higher order
also contain extra independent informations on the distribution function.
The derivation of these
moments requires similar perturbative calculations, and have
been calculated (Bernardeau 1992a and b) when the filter
function is neglected (which turns out to be
equivalent to the limit $n\to-3$).
The results are of the following form
$$\mg\dta^p\md_c\sim S_p\mg\dta^2\md^{p-1}\eqno(71)$$
for the density field and
$$\mg\theta^p\md_c\sim S_{p\theta}\mg\theta^2\md^{p-1}\eqno(72)$$
for the divergence of the velocity field.
The $\Omega$ and $\Lambda$ of the coefficients $S_p$ are very weak but for
the coefficients $S_{p\theta}$ they read
$$S_{p\theta}(\Lambda,\Omega)=[f(\Lambda,\Omega)]^{2-p}\,
S_{p\theta}(\Lambda=0,\Omega=1)\approx {1\over \Omega^{0.6 (p-2)}}
\,S_{p\theta}(\Lambda=0,\Omega=1).\eqno(73)$$
This dependence with the cosmological parameters is expected to
hold even with the application of a filter function (as verified
for $p=3,4$). It implies that
whatever the number of moments you can measure it is impossible to
give any constraint on $\Lambda$. Actually this property comes from the
fact that the divergence of the velocity, at any order in the expansion
(27), always depends on $\Omega$ and $\Lambda$ like $f(\Omega,\Lambda)$,
$\theta^{(p)}(R_0)\propto f(\Omega,\Lambda)$, so that, at large scale,
any statistical indicator can only give a constraint on this combination
of $\Omega$ and $\Lambda$.

\vskip .5 cm
\leftline{4.2. \sl The Zel'dovich approximation}

It has been claimed that the Zel'dovich approximation (Zel'dovich 1970)
was an extremely good and powerful approximation for the large scale
dynamics (\eg, Kofman \etal 1993).
This approximation, based on a linear approximation for the displacement field,
is a simplified dynamics that however reproduces
most of the large-scale features of the gravitational dynamics.
Obviously the results that have been obtained
in the previous section do take into account corrections to the
Zel'dovich approximation when necessary, that is the second or third order
of the displacement field (see Bouchet \etal 1992 for an interesting
discussion on this subject). It is anyway possible to calculate
the large scale behaviors of the density and velocity moments
in the simplified Zel'dovich dynamics. We
obtain for $P(k)\propto k^n$,
$$\eqalign{
S\zel_3&=4-(3+n);\cr
S\zel_{3\theta}&={-1\over \Omega^{0.6}}\left[2-(3+n)\right];\cr
S\zel_4&={272\over 9}-{50\over3}(3+n)+{7\over 3} (3+n)^2;\cr
S\zel_{4\theta}&={1\over \Omega^{1.2}}\left[8-{26\over3}(3+n)+
{7\over 3} (3+n)^2\right]
.\cr}\eqno(74)$$
The $\Omega$ and $\Lambda$ dependences are roughly the same as for
the general results.
As can be seen from these results the Zel'dovich approximation leads
to  an underestimation of the correct values, especially for the
velocity field since for instance for $n=-1$ the predicted skewness and
kurtosis are zero.
The use of a reconstruction method starting with
the velocity field using the Zel'dovich approximation (Nusser \& Dekel 1993,
Gramann 1993)
to constrain $\Omega$ then may lead to a wrong estimation of $\Omega$.

\vskip .5 cm
\leftline{4.3. \sl The biases}

As mentioned previously the observations in galaxy fields provide good
indications in favor of Gaussian initial conditions. However in case
of biases between the galaxy distribution and the matter field,
one can wonder whether the demonstration still holds.
In fact, as pointed out by Fry \& Gazta\~naga (1992), the
scaling between the moments of the galaxy distribution at large scale
is expected
to hold for a broad range of bias schemes. However, neither $S_3$
or $S_4$ can be
used to give constraints on the galaxy linear bias, $b$ (Eq. [3]).
The reason is that
the galaxy field skewness and kurtosis would be not only sensitive
to the linear term $b\,\dta$ but also to the quadratic or cubic corrections.
Following Fry \& Gazta\~naga, if we assume that,
$$\dta_g=b\ \dta+{b_2\over2}\dta^2+{b_3\over 3!}\dta^3+\dots\eqno(75)$$
then we have,
$$\eqalign{
\Sgal_3&={S_3\over b}+3{b_2\over b^2};\cr
\Sgal_4&={S_4\over b^2}+12 {S_3 b_2\over b^3}+4 {b_3\over b^3}+
12 {b_2^2\over b^4}.\cr
}\eqno(76)$$
It shows that any measurement of
high order moments involves a new parameter in the expansion (75), so that
no new constraints on $b$ alone can be given. However a departure between
the measured values of $\Sgal_3$ and $\Sgal_4$ and the theoretical predictions
is the manifestation of biases that is $\dta_g$ is not equal
to $\dta$. For instance neither the skewness measured
in the IRAS galaxy sample by (Bouchet \etal 1993) nor the kurtosis are
in agreement with the theoretical predictions. These are indications
of the existence of biases in the galaxy distribution, although the value
for $b$ defined as the linear bias cannot be inferred from these measurements.

\vskip .5 cm
\leftline{4.4. \sl A universal function to test the gravitational
instability scenario}

Unlike the density field, the velocity divergence
field is thought to be free of unknown
biases. This supposes that it is possible to do a volume weighted
filtering of the field as done by Juszkiewicz \etal (1993b)
in a numerical simulation. For real data, however, it may not be
easy (due for instance to undersampling in voids) but it
is
not the purpose of this paper to test the reliability of these tests on the
present data. The skewness has already been
proposed as a good indicator for the measurement of $\Omega$
and measured in observational data set (Bernardeau \etal 1993).
In fact it turns out that the kurtosis could be  used as well to measure
$\Omega$, giving the possibility to measure $\Omega$ with two independent
methods that are based on the same physical assumptions. Obviously
the two results should be in agreement with each other. In fact
there is a combination of the first three moments of the divergence of the
velocity field that is almost independent of the cosmological parameters,
$$\eqalign{
{S_{4\theta}\over S_{3\theta}^2}={\left(\mg\theta^4\md-
3\mg\theta^2\md^2\right)\mg\theta^2\md
\over \mg\theta^3\md^2}&={{12088\over 441}-{338\over21}(n+3)+{7\over3}(n+3)^2
\over \left[{26\over 7}-(n+3)\right]^2}\cr
}\eqno(77)$$
The values taken by this ratio are given in table 1. The case $n=1$ has not
been given since the velocity skewness vanishes for $n\approx 0.7$ leading
to an unstable ratio. In practice
one should use the index observed in the velocity field rather
than the one of the galaxy clustering since the latter may be affected
by biases. However in the absence of such determination and to illustrate
this result we can use the index observed at 10 $h^{-1}$Mpc in galaxy
catalogues. It equals  approximately
-1 (there are numerous determinations but see for instance
Peacock 1991) so that we expect a ratio of the order of 1.5 (table
1 and Figs. 5-8).
{\sl Then, if the measured large scale velocity flows
fail to reproduce such a number, they cannot originate from gravitational
instabilities with Gaussian initial conditions.}

\vskip .5 cm
\leftline{4.5. \sl Conclusion}

I gave here a series of results for the dependence of the skewness and the
kurtosis of the cosmic fields at large scales with the cosmological
parameters. These results have
been exactly derived using perturbation theory.
For the comparison with observational data however,
numerous effects have to be been taken into
account such as  the redshift space distorsion for the
density field (see Bouchet \etal 1992 for the skewness),
the inhomogeneous Malmquist bias for the velocity field.
Anyway the analysis of the statistical properties of the large--scale
cosmic fields looks promising and it should be
possible to constraint the density
of the universe or to check the validity of the gravitational
instability scenario.

\vskip 1 cm

\leftline{Acknowledgment: }
I thank Hugh Couchman for providing his CDM numerical
simulation and the referee, Enrique Gazta\~naga, for valuable comments
on the manuscript.

 \vfill\eject

\leftline{\bf APPENDIX A: Geometrical properties of the top hat filter
function}
\vskip .5 cm

This appendix is devoted to the derivation of general properties of the
top hat window function, $\WTH(\vx)$. It is defined by
$$\eqalign{
\WTH(\vx)&=1 \hbox{  if  } \vert\vx\vert\le R_0;\cr
\WTH(\vx)&=0 \hbox{  otherwise};\cr}\eqno(A.1)$$
for a scale $R_0$.
The Fourier transform of this function is then given by
$$\WTH(k\,R_0)={3\over (kR_0)^3} \left(\sin(kR_0)-kR_0 \cos(kR_0)\right)
\eqno(A.2)$$
where $k$ is the norm of $\vk$.

\vskip .5 cm
\leftline{A.1. \sl General properties}

Let me define two functions used in the main part:
$$\put=\pu\eqno(A.3)$$
and
$$\qut=\qu\eqno(A.4)$$
Then if one considers two wave vectors $\vk_1$ and $\vk_2$
on which one may have to integrate, a  property of interest
concerning the integrations over
their angular parts, $\d \Om_1$ and $\d \Om_2$, have already
been given by Bernardeau (1993, Appendix B). It reads
$$\eqalign{
{1\over(4\pi)^2}&\int\d \Om_1\d\Om_2\
\qut\WTHd\cr
&=\ {2\over 3}\ \WTH(k_1\,R_0) \WTH(k_2\,R_0).\cr}\eqno(A.5)$$

For the purpose of this paper
we need a few other properties involving other geometrical dependences.

They are the following:

$$\eqalign{
{1\over (4\pi)^2}&\int\d \Om_1\d\Om_2\
\put\WTHd\cr
&=
\WTH(k_1\,R_0)
\left(\WTH(k_2\,R_0)+{1\over 3} k_2\,R_0\,\WTH'(k_2\,R_0)\right)
\cr}\eqno(A.6)$$
and
$$\eqalign{
{1\over (4\pi)^2}&\int\d \Om_1\d\Om_2\
\qut\,
\vert\vk_1+\vk_2\vert R_0\,\dWTHd
\cr
&={2\over 3}\
\left(k_1\,R_0\,\WTH'(k_1\,R_0)\,\WTH(k_2\,R_0)+
\WTH(k_1\,R_0)\,k_2\,R_0\,\WTH'(k_2\,R_0)\right)\cr}\eqno(A.7)$$
and
$$\eqalign{
{1\over (4\pi)^2}&\int\d \Om_1\d\Om_2\
\put\,\vert\vk_1+\vk_2\vert R_0\,\dWTHd
\cr
=&
k_1\,R_0\,WTH'(k_1\,R_0)\,\WTH(k_2\,R_0)
+{1\over 3}k_1\,R_0\,\WTH'(k_1\,R_0)\,k_2\,R_0\,\WTH'(k_2\,R_0)\cr
&-{1\over 3}\WTH(k_1\,R_0)\,k_2^2\,R_0^2\,\WTH(k_2\,R_0).\cr}\eqno(A.8)$$

The expression (A.6) can be derived straightforwardly from the expression
(A.2) of the Fourier transform of the top hat window function.
The integral over the angular part $u={\vk_1.\vk_2/ k_1 k_2}$
can be done by the change of variable, $x=\vert\vk_1+\vk_2\vert\,R_0$, so that
the integral reads,
$$\eqalign{
{1\over (4\pi)^2}&\int\d \Om_1\d\Om_2\
\left[1+{k_1\over k_2}\ {\vk_1.\vk_2\over k_1 k_2}\right]
\WTHd\cr
&=
{3\over2}\int_{\vert\vk_1-\vk_2\vert\,R_0}^{\vert\vk_1+\vk_2\vert\,R_0}
{x\,\d x\over k_1 k_2 R_0^2}\,
\left[1+{k_1\over k_2}\,{x^2/R_0^2-k_1^2-k_2^2\over k_1 k_2}\right]
\,{[\sin x-x\,\cos x]\over x^3}\cr}\eqno(A.9)$$
The integration of (A.9) can be done directly and leads to the expression
(A.6).

The two other properties can be shown using the properties of the
derivative of $\WTH(k)$. Indeed we have
$${1\over 3}\,k\WTH'(k)=-\WTH(k)+{\sin k\over k}.\eqno(A.10)$$
The latter term can be written with a Bessel function,
${\sin k/ k}=\sqrt{\pi/ 2}\,{J_{1/2}(k)/k^{1/2}}$, and
and then decomposed as a sum over products of Bessel functions,
$${\sin\vert\vk_1+\vk_2\vert\over \vert\vk_1+\vk_2\vert}=
\pi\,\sum_{m=0}^{\infty}({1\over 2}+m){J_{1/2+m}(k_1)\over k_1^{1/2}}
{J_{1/2+m}(k_2)\over k_2^{1/2}}\,P_m(-u),\eqno(A.11)$$
where
$$u={\vk_1.\vk_2\over k_1 k_2}.\eqno(A.12)$$
and $P_m$ is the Legendre polynomial of order $m$.
The general properties of the Legendre polynomials then lead
to the relationships,
$$\eqalign{
{1\over 2}\int_{-1}^{1}\d u\, P_m(-u)&=1\ \hbox{  if  }\ m=0, \ \ 0\
\hbox{  otherwise;}\cr
{1\over 2}\int_{-1}^{1}\d u\, u\,P_m(-u)&=-1/3\ \hbox{  if  }\ m=1,\ \  0\
\hbox{  otherwise;}\cr
{1\over 2}\int_{-1}^{1}\d u\, u^2\,
P_m(-u)&=1/3\ \hbox{  if  }\ m=0, \ \ 2/15\ \hbox{  if  }\
m=2, \ \ 0\ \hbox{  otherwise.}\cr}\eqno(A.13)$$

Using these relationships we obtain,
$$\eqalignno{
{1\over 2}\int_{-1}^{1}\d u[1-u^2]{\sin\vert\vk_1+\vk_2\vert\over
\vert \vk_1+\vk_2\vert}=&
{2\over3}{\sin k_1\over k_1}{\sin k_2\over k_2}-{\pi\over 3}
{J_{5/2}(k_1)\over k_1^{1/2}}{J_{5/2}(k_2)\over k_2^{1/2}}&(A.14)\cr
{1\over 2}\int_{-1}^{1}\d u\left[1+{k_1\over k_2}u\right]
{\sin\vert\vk_1+\vk_2\vert\over
\vert \vk_1+\vk_2\vert}=&
{\sin k_1\over k_1}{\sin k_2\over k_2}-{\pi\over 2}{k_1\over k_2}
{J_{3/2}(k_1)\over k_1^{1/2}}{J_{3/2}(k_2)\over k_2^{1/2}}&(A.15)\cr}
$$

It can also be noticed that ,
$$\eqalignno{
J_{3/2}(k)&=k^{3/2}\,\WTH(k)\,\sqrt{2\over 3\pi}&(A.16),\cr
J_{5/2}(k)&=-k^{3/2}\,\WTH'(k)\,\sqrt{2\over 3\pi}&(A.17),\cr}$$
so that
$$\eqalignno{
{1\over 2}\int_{-1}^{1}\d u[1-u^2]&{\sin\vert\vk_1+\vk_2\vert\over
\vert \vk_1+\vk_2\vert}=\cr
&{2\over3}\left({\sin k_1\over k_1}{\sin k_2\over k_2}-{1\over 9}
k_1\WTH'(k_1)\,k_2\WTH'(k_2)\right)&(A.18)\cr
{1\over 2}\int_{-1}^{1}\d u\left[1+{k_1\over k_2}u\right]
&{\sin\vert\vk_1+\vk_2\vert\over
\vert \vk_1+\vk_2\vert}=
{\sin k_1\over k_1}{\sin k_2\over k_2}-{1\over9}k_1^2\,\WTH(k_1)\WTH(k_2).
&(A.19)\cr}$$
The relations (A.7) and (A.8) can then be deduced from (A.10), (A.18)
and (A.19).

\vskip .5 cm
\leftline{A.2. \sl Applications}

As an application of the previous results I give the value of few integrals
of interest for the large scale cosmic fields. Let me define $\sigma^2(R_0)$
by
$$\sigma^2(R_0)=\int{\d^3\vk\over (2\pi)^3} P(k)\,\WTH^2(k\,R_0)\eqno(A.20)$$
where $P(k)$ is the power spectrum.
We can then notice that
$$R_0\,{\d\sigma^2(R_0)\over \d R_0}=
2\int{\d^3\vk\over (2\pi)^3} P(k)\,k\,R_0\,\WTH(k\,R_0)\WTH'(k\,R_0).
\eqno(A.21)$$

The second derivative can also be calculated,
$$\eqalign{
R_0\,{\d\over \d R_0}&\left[R_0\,{\d\sigma^2(R_0)\over \d R_0}\right]=
2\int{\d^3\vk\over (2\pi)^3} P(k)\,\left[k\,R_0\,\WTH'(k\,R_0)\right]^2\cr
&+2\int{\d^3\vk\over (2\pi)^3} P(k)\,\left[k\,R_0\right]^2\,
\WTH(k\,R_0)\WTH''(k\,R_0)\cr
&+2\int{\d^3\vk\over (2\pi)^3} P(k)\,k\,R_0\,\WTH(k\,R_0)\WTH'(k\,R_0).\cr}
\eqno(A.22)$$

The expression (A.22) can be simplified by the following property,
$$k^2\,\WTH''(k)=-4k\,\WTH'(k)-k^2\,\WTH(k),\eqno(A.23)$$
so that
$$\eqalign{
R_0\,{\d\over \d R_0}\left[R_0\,{\d\sigma^2(R_0)\over \d R_0}\right]=&
2\int{\d^3\vk\over (2\pi)^3} P(k)\,\left[k\,R_0\,\WTH'(k\,R_0)\right]^2\cr
&+3\,R_0\,{\d\sigma^2(R_0)\over \d R_0}+
2\,\overline{k^2}\,\sigma^2(R_0)\cr}\eqno(A.24)$$
where
$$\overline{k^2}={1\over \sigma^2(R_0)}\int{\d^3\vk\over (2\pi)^3} k^2\,P(k)
\,\WTH^2(k\,R_0)\eqno(A.25)$$

I consider now the integrals,
$$\eqalign{
I_1(R_0)=\intdc\,&\mP(\vk_1,\vk_2)\,P(k_1)\,P(k_2)\cr
&\times\WTHd\WTH(k_1\,R_0)\WTH(k_2\,R_0)\cr
I_2(R_0)=\intdc\,&\mQ(\vk_1,\vk_2)\,P(k_1)\,P(k_2)\cr
&\times\WTHd\WTH(k_1\,R_0)\WTH(k_2\,R_0)\cr}\eqno(A.26)$$
The relationships (A.3) and (A.4) give,
$$\eqalign{
I_1(R_0)&=\sigma^4(R_0)\left(1+{1\over6}{R_0\over\sigma^2(R_0)}\,
{\d\sigma^2(R_0)\over \d R_0}\right)\cr
I_2(R_0)&={2\over3}\,\sigma^4(R_0)\cr}\eqno(A.27)$$

The knowledge of integrals involving three wave vectors are also
required. To ease the presentation I introduce simplified notations,
$$
\eqalign{
W_i&=\WTH(k_i\,R_0),\
W_{ij}=\WTH\left(\vert\vk_i+\vk_j\vert\ R_0\right),\ \cr
W_{ijk}&=\WTH\left(\vert\vk_i+\vk_j+\vk_k\vert\ R_0\right).\cr
\mP_{i,j}&=\mP(\vk_i,\vk_j),\ \mP_{ij,k}=\mP(\vk_i+\vk_j,\vk_k),\
\mP_{i,jk}=\mP(\vk_i,\vk_j+\vk_k).\cr}$$
and similar notations for $\mQ(\vk_i,\vk_j)$. Then we have the results,
$$\eqalign{
\inttc&\,P(k_1)\,P(k_2)\,P(k_3)\,W_1\,W_{12}\,W_{23}\,W_3\,
\mP_{1,2}\,\mP_{2,3}\cr
&=\sigma^6(R_0)\left(1+{1\over 3} \gm_1+{1\over 36}\gm_1^2\right)\cr
\inttc&\,P(k_1)\,P(k_2)\,P(k_3)\,W_1\,W_{12}\,W_{23}\,W_3\,
\mP_{1,2}\,\mP_{3,2}\cr
&=\sigma^6(R_0)\left(1+{1\over 2} \gm_1+{1\over 18}\gm_1^2+
{1\over 18}\gm_2+{1\over 9}\overline{k^2}\right)\cr
\inttc&\,P(k_1)\,P(k_2)\,P(k_3)\,W_1\,W_{12}\,W_{23}\,W_3\,
\mP_{2,1}\,\mP_{2,3}\cr
&=\sigma^6(R_0)\left(1+{1\over 3} \gm_1+{1\over 36}\gm_1^2\right)\cr
\inttc&\,P(k_1)\,P(k_2)\,P(k_3)\,W_1\,W_{12}\,W_{23}\,W_3\,
\mP_{1,2}\,\mQ_{2,3}\cr
&={2\over3}\sigma^6(R_0)\left(1+{1\over 6} \gm_1\right)\cr
\inttc&\,P(k_1)\,P(k_2)\,P(k_3)\,W_1\,W_{12}\,W_{23}\,W_3\,
\mP_{2,1}\,\mQ_{2,3}\cr
&={2\over3}\sigma^6(R_0)\left(1+{1\over 6} \gm_1\right)\cr
\inttc&\,P(k_1)\,P(k_2)\,P(k_3)\,W_1\,W_{12}\,W_{23}\,W_3\,
\mQ_{1,2}\,\mQ_{2,3}={4\over 9}\sigma^6(R_0),\cr
}\eqno(A.28)$$

where
$$\eqalign{
\gm_1={\d\log[\sigma^2(R_0)]\over \d\log R_0},\ \
\gm_2={\d^2\log[\sigma^2(R_0)]\over \d\log^2 R_0}.\cr}\eqno(A.29)$$

Another series of useful integrals are given by,
$$\eqalign{
\inttc&\,P(k_1)\,P(k_2)\,P(k_3)\,W_1\,W_{2}\,W_{3}\,W_{123}\,
\mP_{12,3}\,\mP_{1,2}\cr
&=\sigma^6(R_0)\left(1+{1\over 3} \gm_1+{1\over 36}\gm_1^2\right)\cr
\inttc&\,P(k_1)\,P(k_2)\,P(k_3)\,W_1\,W_{2}\,W_{3}\,W_{123}\,
\mP_{3,12}\,\mP_{1,2}\cr
&=\sigma^6(R_0)\left(1+{1\over 3} \gm_1+{1\over 36}\gm_1^2-
{1\over 9}\overline{k^2}\right)\cr
\inttc&\,P(k_1)\,P(k_2)\,P(k_3)\,W_1\,W_{2}\,W_{3}\,W_{123}\,
\mP_{12,3}\,\mQ_{1,2}\cr
&={2\over3}\sigma^6(R_0)\left(1+{1\over 6} \gm_1\right)\cr
\inttc&\,P(k_1)\,P(k_2)\,P(k_3)\,W_1\,W_{2}\,W_{3}\,W_{123}\,
\mP_{3,12}\,\mQ_{1,2}\cr
&={2\over3}\sigma^6(R_0)\left(1+{1\over 3} \gm_1\right)\cr
\inttc&\,P(k_1)\,P(k_2)\,P(k_3)\,W_1\,W_{2}\,W_{3}\,W_{123}\,
\mQ_{12,3}\,\mP_{1,2}\cr
&={2\over3}\sigma^6(R_0)\left(1+{1\over 6} \gm_1\right)\cr
\inttc&\,P(k_1)\,P(k_2)\,P(k_3)\,W_1\,W_{2}\,W_{3}\,W_{123}\,
\mQ_{12,3}\,\mQ_{1,2}={4\over9}\sigma^6(R_0).\cr
}\eqno(A.30)$$

 \vfill\eject

\centerline{\bf REFERENCES}
{\parindent=0pt
\apjref Bernardeau, F. 1992a; ApJ; 392; 1;
\apjref Bernardeau, F. 1992b; ApJ; 390; L61;
\prepref Bernardeau, F. 1993; ApJ in press;
\prepref Bernardeau, F., Juszkiewicz, R., Dekel, A. \& Bouchet, F. ;
in preparation;
\apjref Bertschinger, E., Dekel, A., Faber, S.M., Dressler, A., Burstein, D.,
1991; ApJ; 364; 370;
\apjref Bond, J.R., Cole, S., Efstathiou, G., \& Kaiser, N. 1991; ApJ;
379; 440;
\apjref Bouchet, F. \& Hernquist, L. 1992; ApJ; 400; 25;
\apjref Bouchet, F., Juszkiewicz, R., Colombi, S., \& Pellat, R. 1992; ApJ;
394; L5;
\apjref Bouchet, F., Strauss, M.A., Davis, M., Fisher, K.B., Yahil, A.
\& Huchra, J.P. 1993; ApJ; 417; 36;
\apjref Couchman, H.M.P., 1991; ApJ; 368; L23;
\apjref Dekel, A., Bertschinger, E., Faber, S.M., 1990; ApJ.; 364;349;
\apjref Efstathiou, G. Frenk, C.S., White, S.D.M., Davis, M., 1988; MNRAS;
235; 715;
\apjref Fry, J.N., 1984; ApJ; 279; 499;
\apjref Fry, J.N. \& Gazta\~naga, E., 1992; ApJ; 413; 447;
\apjref Goroff, M.H., Grinstein, B., Rey, S.-J. \& Wise, M.B. 1986; ApJ;
311; 6;
\apjref Gramann, M. 1993; ApJ; 405; 449;
\apjref Hamilton, A.J.S., Kumar, P., Lu, E., \& Matthews, A. 1991; ApJ;
374; L1;
\apjref Juszkiewicz, R., Bouchet, F. \& Colombi, S., 1993a;
ApJ; 419; L9;
\prepref Juszkiewicz, R., Weinberg, D. H., Amsterdamski, P., Chodorowski, M.
\& Bouchet, F. 1993b; IASSNS-AST 93/50 preprint;
\apjref Kaiser, N. \& Lahav, O. 1989; MNRAS; 237; 129;
\prepref Kofman, L., Bertschinger, E., Gelb, M.J., Nusser, A., Dekel, A. 1993;
CITA 93/13 preprint;
\apjref Lahav, O., Lilje, P.B., Primack, J.R. \& Rees, M.J. 1991; MNRAS;
251; 128;
\apjref Lynden-Bell, D., Faber, S.M., Burstein, D., Davies, R.L.,
Dressler, A., Terlevich, R.J. \& Wegner, G. 1988; ApJ; 326; 19;
\apjref Nusser, A. \& Dekel, A. 1993; ApJ; 405; 437;
\apjref Nusser, A., Dekel, A., Bertschinger, E. \& Blumenthal, G. 1991; ApJ;
379; 6;
\apjref Peacock, J.A. 1991; MNRAS; 253; 1p;
\apjref Peacock, J.A. 1992; MNRAS; 258; 581;
\bookref Peebles, P.J.E. 1980;  The Large Scale Structure of the Universe;
Princeton University Press, Princeton, N.J., USA;
\apjref Rowan-Robinson, M., Longmore, A., Saunders, W., Ellis, R.S., Frenk,
C.S., Parry, I., Xiaoyang, X., Allington-Smith, J.R.,
Efstathiou, G. \& Kaiser, N. 1990; MNRAS; 247; 1;
\apjref Shaya, E.J., Tully, R.B. \& Pierce, M.J. 1992; ApJ; 391; 16;
\prepref Strauss, M. \& Davis, M. 1988; In {\sl Large-scale motions in the
Universe,} eds Rubin, V.C. \& Coyne, G., Princeton University Press,
Princeton, NJ.;
\prepref Yahil, A. 1991; In {Proc. Particle Astrophysics, 10th Moriond
Astrophysics Meeting}, eds Alimi, J.M., Blanchard, A., Bouquet, A.,
Martin de Volnay, F. \& Tr\^an Thanh V\^an, J. Editions Fronti\`eres,
Gif-sur-Yvette.;
\apjref Zel'dovich, Ya.B. 1970; A \& A; 5; 84;
}

 \vfill\eject

\centerline{\bf FIGURE CAPTIONS}
\noindent
\vskip .5cm
{\sl Fig. 1 :}
Variation of $D_2$ (Eq. [39]) and $E_2$ (Eq. [43]) as a function of
$\Omega$. Two cosmological hypothesis are considered: the solid line
if for $\Lambda=0$ and the dashed line for a flat universe,
$\Omega+\Lambda/3H^2=1$.
\vskip .5 cm

{\sl Fig. 2 :}
Variation of $D_3$ (Eq. [40]) and $E_3$ (Eq. [44]) as a function of
$\Omega$. The line symbols are the same as in Fig. 1.
\vskip .5 cm

{\sl Fig. 3 :}
Large--scale behavior of the skewness of the density field (top) and of
the divergence of the density field (bottom) as a function of $\Omega$.
The lines correspond to various index of the
power spectrum index: from top to bottom $n=-3$, $-2$, $-1$, $0$, $+1$
in each panel and the
solid lines are for $\Lambda=0$ and the dashed lines for
$\Omega+\Lambda/3H^2=1$.
\vskip .5 cm

{\sl Fig. 4 :}
The same as Fig. 3 but for the kurtosis of the density and velocity fields.
In the latter case the $n=1$ case is not presented since
$S_{4\theta}$ is close to zero.
\vskip .5 cm

{\sl Fig. 5 :}
The same as Fig. 3 but for the combination of the first three
moments that is shown to have a very weak dependence with
the cosmological parameters for the two fields.
\vskip .5 cm

{\sl Fig. 6 :}
The moments as a function of the power spectrum index. The left panels
are for the density field and the right for the divergence of the velocity
field. The top panels give the skewness, the middle give the kurtosis and
the bottom present a combination of these moments (Eq. [77]). The solid lines
correspond to $\Omega=1$, $\Lambda=0$, the long dashed line to $\Omega=0.3$,
$\Lambda=0$ and the short dashed lines to $\Omega=0.3$, $\Lambda/3H^2=0.7$.
\vskip .5 cm

{\sl Fig. 7 :}
The moments as a function of scale in case of a CDM power spectrum.
The top panels give the local index $n$ (Eq. [70], thick lines)
and $\gamma_2\ (\equiv -\d n/\d\log(R_0))$ (thin lines)
as a function of the smoothing radius.
The functions $S_3$ and $S_4$ are calculated according to Eqs. [56-59]
(thick lines) and thin lines correspond to the approximation
(67-68) when the power spectrum is locally approximated by a power law
of index $n$. The other symbols are the same as in Fig. 6.
The perturbation theory is thought to be valid for smoothing radius
greater than $\sim 10 h^{-1}$ Mpc ($h=H_0/100\ $km s$^{-1}$Mpc$^{-1}$).
The squares
with the error bars are the measurements obtained in a CDM numerical
simulation with $\Omega=1$ presented in table 1 (and for $a/a_0=1$).

\vskip .5 cm

{\sl Fig. 8 :}
The moments as a function of scale in case of the power spectrum
proposed by Peacock (1991) to fit the observed galaxy correlation
function. Symbols are the same as in Figs. 6-7.
\vskip .5 cm

\end